\documentclass[prd, twocolumn, nofootinbib, showkeys, preprintnumbers]{revtex4}
\usepackage{graphicx}
\usepackage{dcolumn}
\usepackage{epsfig} 

\begin{document} 

\title{Testing Cosmology with Cosmic Sound Waves}

\author{Pier Stefano Corasaniti$^1$ and Alessandro Melchiorri$^{2,3}$} 
\affiliation{
$^1$LUTH, Observatoire de Paris, CNRS UMR 8102, Universit\'e Paris Diderot, 5 Place Jules Janssen, 92195 Meudon Cedex, France\\
$^2$Dipartimento di Fisica e Sezione INFN, Universita' degli Studi di
Roma ``La Sapienza'', Ple Aldo Moro 5,00185, Rome, Italy\\
$^3$CERN, Theory Division, CH-1211 Geneva 23, Switzerland}

\begin{abstract}
WMAP observations have accurately determined the position of the
first two peaks and dips in the CMB temperature power spectrum. These
encode information on the ratio of the distance to the last scattering
surface to the sound horizon at decoupling. However pre-recombination
processes can contaminate this distance information. In order to assess the 
amplitude of these effects we use the WMAP data and evaluate the relative 
differences of the CMB peaks
and dips multipoles. We find that the position of the first
peak is largely displaced with the respect to the expected position of the sound
horizon scale at decoupling. In contrast the relative spacings 
of the higher extrema are statistically consistent with 
those expected from perfect harmonic oscillations. 
This provides evidence for a scale dependent phase shift 
of the CMB oscillations which is caused by gravitational driving
forces affecting the propagation of sound waves before
recombination. By accounting for these effects we have 
performed a MCMC likelihood analysis of the location of WMAP extrema to
constrain in combination with recent BAO data a constant dark energy 
equation of state parameter $w$. For a flat universe we find a strong 
$2\sigma$ upper limit $w<-1.10$, and including the HST prior we obtain 
$w<-1.14$, which are only marginally consistent with limits derived
from the supernova SNLS sample. On the other hand we infer larger
limits for non-flat cosmologies. From the full CMB likelihood analysis we also estimate the
values of the shift parameter $R$ and the multipole $l_a$ of the 
acoustic horizon at decoupling for several cosmologies to test their 
dependence on model assumptions. Although the analysis of the full CMB
spectra should be always preferred, using the position of  
the CMB peaks and dips provide a simple and consistent 
method for combining CMB constraints with other datasets.
\end{abstract} 

\date{\today}
\preprint{CERN-PH-TH/2007-241}
\maketitle 

\section{Introduction}\label{intro}

Cosmic Microwave Background (CMB) observations have
provided crucial insights into the origin and evolution 
of present structures in the universe \cite{CMBobs1,CMBobs2,Hinshaw}.
Physical processes occurred before, during 
and after recombination have left distinctive signatures on the CMB. 
The most prominent feature is a sequence of peaks and dips in the
anisotropy power spectrum, the remnant imprints of acoustic waves 
propagating in the primordial photon-baryon plasma at the
time of decoupling \cite{CMBTh1,CMBTh2,CMBTh3}. 
This oscillatory pattern carries specific information 
on several cosmological parameters \cite{PeaksTh}. As an example 
the angular scale at which these oscillations are observed 
provides a distance measurement of the last scattering surface
to the sound horizon at decoupling, hence a clean test of 
cosmic curvature \cite{Ktest}.

WMAP observations have accurately detected the peak structure 
of the CMB power spectrum. These data have constrained the geometry of 
the universe to be nearly flat and have precisely 
determined other cosmological parameters \cite{Spergel}. 
On the other hand constraints on dark energy are less stringent, this
is because its late time effects leave a weaker imprint of the CMB
which is diluted by degeneracies with other
parameters. Indeed other cosmological tests can be more sensitive to
the signature of dark energy, nonetheless they 
still require additional information
from CMB to break the parameter degeneracies. As an example
CMB constraints are usually combined with those from SN Ia luminosity 
distance data. Alternatively the CMB can be used in combination
with measurements of the baryon acoustic oscillations (BAO) in 
the galaxy power spectrum \cite{Eisenstein}.
In fact the same acoustic signature present in the CMB is also 
imprinted in the large scale distribution of galaxy, thus providing
a complementary probe of cosmic distances at lower redshifts.

A likelihood analysis of the CMB spectra is certainly the more robust
approach to implement CMB constraints with those from other datasets.
This can be very time consuming, henceforth one can try 
to compress the CMB information in few measurable and easily
computable quantities. Recent literature 
has focused on the use of the shift parameter $R$, 
and the multipole of the acoustic scale at decoupling $l_a$ 
\cite{Elgaroy,Wang}. 
However these quantities are not directly measured by CMB
observations, they are inferred as secondary parameters 
from the cosmological constraints obtained from the full 
CMB likelihood analysis. Consequently their use as {\it data} 
can potentially lead to results which suffer of model dependencies as well
as prior parameter assumptions made in the analysis 
from which the values of $R$ ($l_a$) have been inferred in the first place. 
In contrast the multipole location of the CMB extrema can be directly 
determined from the observed temperature power spectrum through 
model-independent curve fitting. These measurements can then be used
to constrain cosmological parameters provide that pre-recombination corrections
are properly taken into account.

In this paper we analyse in detail the cosmological information
encoded in the position of the CMB extrema as measured by
WMAP. Our aim is to provide a simple and unbiased method for
incorporating CMB constraints into other datasets 
which is alternative to that of using $R$ and/or $l_a$ \cite{Elgaroy,Wang}.  
Firstly we estimate the amplitude of pre-recombination
mechanisms that can displace the location of the CMB extrema with the 
respect to the angular scale of the sound horizon at decoupling. In particular
we show that the WMAP location of the first peak is strongly affected
by such mechanisms, while the displacements induced on the 
higher peaks and dips are smaller. 
By accounting for these effects we perform a cosmological parameter 
analysis and infer constraints on dark energy under different 
prior assumptions, including the cosmic curvature. 
We then combine these results with
measurements of BAO from SDSS and 2dF data \cite{Percival}, and
confront the inferred constraints with those obtained
using SN Ia data from the Supernova Legagy Survey \cite{Astier}.
Finally we test for potential model dependencies
of $R$ (and $l_a$) by performing a full likelihood analysis of the
WMAP spectra for different sets of cosmological parameters.

The paper is organized as follows: in Section~\ref{Peaks} we review
the physics of the CMB acoustic oscillations. In Section~\ref{WMAP} 
we discuss the relative shifts of the multipoles of the 
WMAP peaks and dips. In Section~\ref{Cosmo} we present the 
results of the cosmological parameter inference using the location of
the CMB extrema in combination with BAO. In Section~\ref{sndata} we 
confront the results with the SN Ia likelihood analysis from the
SNLS sample. We discuss the results on
the shift parameter in Section~\ref{shift} and present our
conclusions in Section~\ref{Conclusions}.

\section{CMB Acoustic Oscillations}\label{Peaks}

The onset of acoustic waves on the sub-horizon scales of 
the tightly coupled photon-baryon plasma before
recombination is natural consequence of photon pressure resisting gravitational 
collapse. The properties of these oscillations
depends both on the background expansion and the evolution of the
gravitational potentials associated with the perturbations present
in the system. In the following we will briefly review the basic processes
which affect the propagation of these waves before
decoupling. Interested readers will find more detailed discussions
in \cite{CMBTh3,PeaksTh}. Let consider the photon temperature
fluctuation $\Theta_0\equiv \Delta T$ (monopole), following 
Hu and Sugiyama \cite{CMBTh3} its evolution is described by
\begin{equation}
\ddot{\Theta}_0+\frac{\dot{R}}{1+R}\dot{\Theta}_0+k^2 c_s^2\Theta_0=F(\eta),\label{monoeq}
\end{equation}
where the dot is the derivative with respect to conformal time, $R=3 \rho_b/4\rho_\gamma$ is 
the baryon-to-photon ratio, $k$ is the wavenumber, $c_s=c/\sqrt{3(1+R)}$
is the sound speed of the system with $c$ the speed of light. The source term
\begin{equation}
F=-\ddot{\Phi}-\frac{\dot{R}}{1+R}\dot{\Phi}-k^2 \frac{\Psi}{3},\label{source}
\end{equation}
represents a driving force, where $\Phi$ and $\Psi$ are the gauge-invariant metric perturbations
respectively.

It is easy to see from Eq.~(\ref{monoeq}) that the homogeneous 
equation ($F=0$) admits oscillating solutions of the form,
\begin{equation}
\Theta^{\rm hom}_0(\eta) =A_1\cos{kr_s(\eta)} 
    +\frac{A_2}{k}\sin{k r_s(\eta)}
\end{equation} 
where $A_1$ and $A_2$ are set by the initial conditions and
$r_s(\eta)=\int_0^\eta c_s(\eta')d\eta'$ is the sound horizon at time
$\eta$. At time of decoupling $\eta_*$, the positive and negative
extrema of these oscillations appear as a series of peaks in the 
anisotropy power spectrum. Their location in the multipole space 
is a multiple integer of the inverse of the angle subtended by the sound
horizon scale at decoupling, namely
$l_m^{peak}=m\,l_a$ with $m=1,2,...$ and
\begin{equation}
l_a = \pi \frac{r_K(z_*)}{r_s(z_*)}, \label{la}
\end{equation} 
where $z_*$ is the recombination redshift and $r(z)$ 
the comoving distance to $z$,
\begin{equation}
r_K(z)=\frac{c}{H_0} \frac{1}{\sqrt{|\Omega_K|}}
f(\sqrt{|\Omega_K|}I(z)),
\end{equation}
with $H_0$ the Hubble constant, $|\Omega_K|=-K/H_0^2$ 
with $K$ the constant curvature, $f(x)=\sin(x),\sinh(x),x$ 
for $K>0,<0$ and $=0$ respectively, and $I(z)=\int_0^z dz' H_0/H(z')$.

Scales for which the monopole vanishes also contribute to
anisotropy power spectrum. In such a case the signal comes
from the non-vanishing photon velocity $\Theta_1$ (dipole) 
which oscillates with a phase shifted by $\pi/2$ with
the respect to the monopole \cite{CMBTh3}. Therefore 
photons coming from these regions are responsible for
a series of troughs in the anisotropy power
spectrum at multipoles $l_n^{dip}=n\,l_a$ with $n=m+1/2$. 

The full solution to Eq.~(\ref{monoeq}) at decoupling reads as \cite{HS}: 
\begin{widetext}
\begin{equation}
\Theta_0(\eta_*) = \Theta^{\rm hom}_0(\eta_*)+\frac{A_3}{k}\int_0^{\eta_*} d\eta'
    [1+R(\eta')]^{3/4} sin{[k r_s(\eta_*) - k r_s(\eta')]}F(\eta'),\\
\label{fullsol}
\end{equation}
\end{widetext}
where $A_3$ is set by the initial conditions. As we can see from
Eq.~(\ref{fullsol}) including the driving force $F$ induces a scale 
dependent phase shift of the acoustic oscillations, 
which is primarily caused by the time variation of the gravitational 
potential $\Phi$. In fact perturbations on scales which enter 
the horizon at the matter-radiation equality experience 
a variation of the expansion rate which causes a time evolution
of the associated gravitational potentials. 
This mechanism is dominant on the large scales and is
responsible for the so called early Integrated Sachs-Wolfe (ISW)
effect \cite{ISW}. 
The overall effect is to displace the acoustic oscillations with 
the respect to the pure harmonic series. For a spectrum of adiabatic
perturbations we may expect this displacement to become negligible 
on higher harmonics since the gravitational potentials decay as
$\Phi\propto(k\eta)^{-2}$ on scales well 
inside the horizon. This is not
the case if active perturbations were present on such scales
before the epoch of decoupling.

In order to account for these pre-recombination effects a realistic modeling of
the multipole position of the CMB maxima and minima is given by \cite{Fukugita} 
\begin{equation}
l_m = l_a (m-\varphi_m),\label{lm}
\end{equation}
where $m=1,2,..$ for peaks, and $m=3/2,5/2,..$ for dips;
$\varphi_m$ parametrizes the displacement caused by the driving
force. Because of the scale dependent nature of the driving
effect discussed above, it is convenient to decompose the correction
term as $\varphi_m=\bar{\varphi}+\delta\varphi_m$, where
$\bar{\varphi}\equiv\varphi_1$ 
is the overall shift of the first peak with respect to the sound
horizon, and $\delta\varphi_m$ is the shift of the m-th extrema 
relative to the first peak \cite{Doran}.

It is worth noticing that while the position of the CMB extrema
depends through $l_a$ on the geometry and late time expansion of the
universe, their relative spacing depends through $\varphi_m$ only on
pre-recombination physics. 
  
\section{Phase shift of WMAP peaks and dips}\label{WMAP}

WMAP observations have provided an accurate determination of
the CMB power spectrum. The multipoles of the CMB extrema have been
inferred using a functional fit to the uncorrelated band powers as
described in \cite{Page}. Hinshaw et al. \cite{Hinshaw}
have applied this method to the WMAP-3yr data and found the position
of the first two peaks and dips to be at $l_1=220.8\pm0.7$,
$l_{3/2}=412.4\pm1.9$, $l_2=530.9\pm3.8$ and $l_{5/2}=675.2\pm11.1$
respectively. 

We want to determine whether these measurements provide any
evidence for driving effects affecting the acoustic oscillations. 
In order to do so we evaluate the relative spacings between the WMAP
measured $m$-th and $m'$-th extrema,
\begin{equation}
\Delta_{m,m'}=\frac{l_{m'}}{l_m}-1,
\end{equation}
and the propagated errors $\sigma_{\Delta_{m,m'}}$.
\begin{figure}[t]
\includegraphics[scale=0.42]{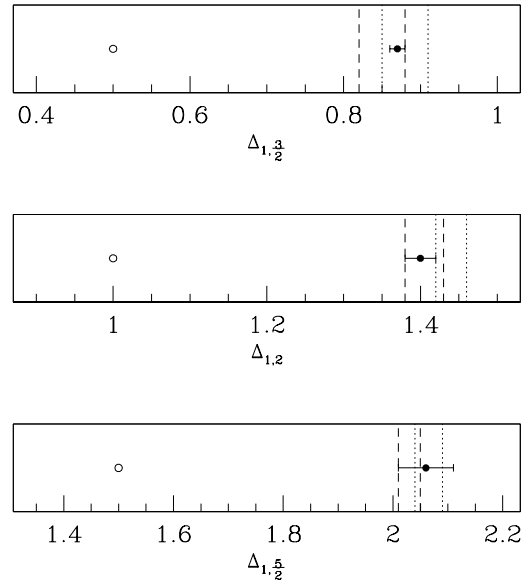}
\caption{WMAP spacings of $l_{3/2}$, $l_2$ and $l_{5/2}$
  relative to $l_1$ (black solid circles) and propagated
  errors. The values expected from the harmonic series 
are $\Delta_{1,3/2}=1/2$,
  $\Delta_{1,2}=1$ and $\Delta_{1,5/2}=3/2$ (open circles). Vertical dashed
  lines delimit the expected interval of variation of the relative
  spacings obtained by 
 including the shift corrections as parametrized in
  \cite{Doran} and evaluated over a conservative range of cosmological
parameter values (see text). The dotted vertical lines include the
  effect of three massless neutrinos.}
\label{fig1}
\end{figure} 

Let first consider the spacings relative to the location of
the first peak. We find $\Delta_{1,3/2}=0.87\pm0.01$, 
$\Delta_{1,2}=1.40\pm0.02$ and $\Delta_{1,5/2}=2.06\pm0.05$
respectively. These estimates are shown in Figure~\ref{fig1} 
(black solid circles), where we also plot the relative spacings
as expected from a sequence of perfect acoustic oscillations 
(open circles). It is evident that the WMAP
inferred values of $\Delta_{1,m}$ lie many sigmas away 
from those expected from the harmonic series. 
This provides clear evidence that the position of the first peak is 
largely affected by the driving force at decoupling. Such a large
displacement is most likely caused by the early ISW, although
an additional contribution from isocurvature fluctuations \cite{Keskitalo} or 
active gravitational potentials \cite{Magueijo} cannot be excluded.

\begin{figure}[t]
\includegraphics[scale=0.42]{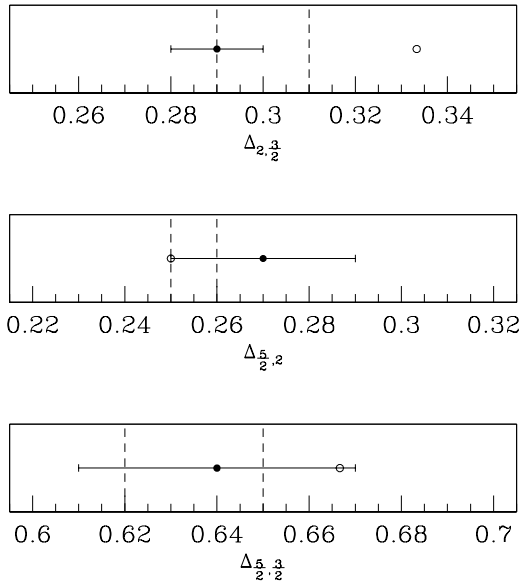}
\caption{As in Figure~\ref{fig1} for $l_{3/2}$, $l_2$ and $l_{5/2}$
  relative spacings. The harmonic series values are $\Delta_{3/2,2}=1/3$, 
  $\Delta_{2,5/2}=1/4$ and $\Delta_{3/2,5/2}=2/3$.}
\label{fig2}
\end{figure}

Let focus now on the displacement of the second peak relative to the first
 one, since $\Delta_{1,2}>1$ it follows that 
$\bar{\varphi}>\delta\varphi_2$. This implies that the overall shift 
of $l_1$ with the respect to $l_a$ is larger than the shift of $l_2$ 
relative to $l_1$. As discussed in the previous section this 
is consistent with having the gravitational potentials inside the
sound horizon scaling as $\Phi\propto(k\eta)^{-2}$, thus
inducing a weaker driving force. 
This can be seen more clearly in Figure~\ref{fig2} 
where we plot $\Delta_{3/2,2}$, $\Delta_{2,5/2}$ 
and $\Delta_{3/2,5/2}$.

Apart $\Delta_{2,3/2}=0.29\pm0.01$, whose value suggests 
the presence of a non-negligible driving effect still on the 
scale of the first dip, we may notice that all other spacings are 
statistically consistent with the prediction of the harmonic series. 

Therefore these results suggest the existence of a scale 
dependent phase shift of the CMB acoustic oscillations. 
The effect is larger on the scale of the first acoustic peak, 
while it is weaker for the higher harmonics. 
The upcoming Planck mission will map
more accurately the location of the higher peaks and dips
and provide a cleaner detection of this shift.
  
Indeed driving effects are well accounted for by the CMB theory 
as incorporated in standard Boltzmann codes \cite{Cmbcode}. For instance
a standard adiabatic spectrum of initial density perturbations leads to phase 
shifts which are consistent with those we have inferred here.
To show this we have used the fitting formulas provided in
\cite{Doran} for adiabatic models which parametrize $\varphi_m$ in terms 
of the total matter density $\Omega_m h^2$, the baryon density $\Omega_b
h^2$, the dark energy density at decoupling $\Omega_{DE}^{dec}$ 
and the scalar spectral index $n_s$. 
Assuming $\Omega_{DE}^{dec}=0$
we evaluate these formulas
over the following range of parameter values, 
$0.08<\Omega_m h^2<0.11$, $0.020<\Omega_b
h^2<0.024$, $0.92<n_s<1.1$ and infer the corresponding intervals
for the relative spacings $\Delta_{m,m'}$. These are drawn in Figure~\ref{fig1} and
\ref{fig2} as vertical dashed lines. It can be seen that these intervals are statistically
consistent with the measured spacings. Including the contribution of three massless neutrinos 
(dotted vertical lines) slightly shifts the $\Delta_{1,m}$ intervals further from the expected
values of the perfect harmonic oscillator.
This is because the presence of relativistic neutrinos 
extend the radiation era and therefore leads to a more effective early ISW
effect on the large scales. In contrast we find no differences for
the intervals of the other peaks and dips spacings.

\begin{figure*}
\centerline{
\includegraphics[scale=0.7,angle=90]{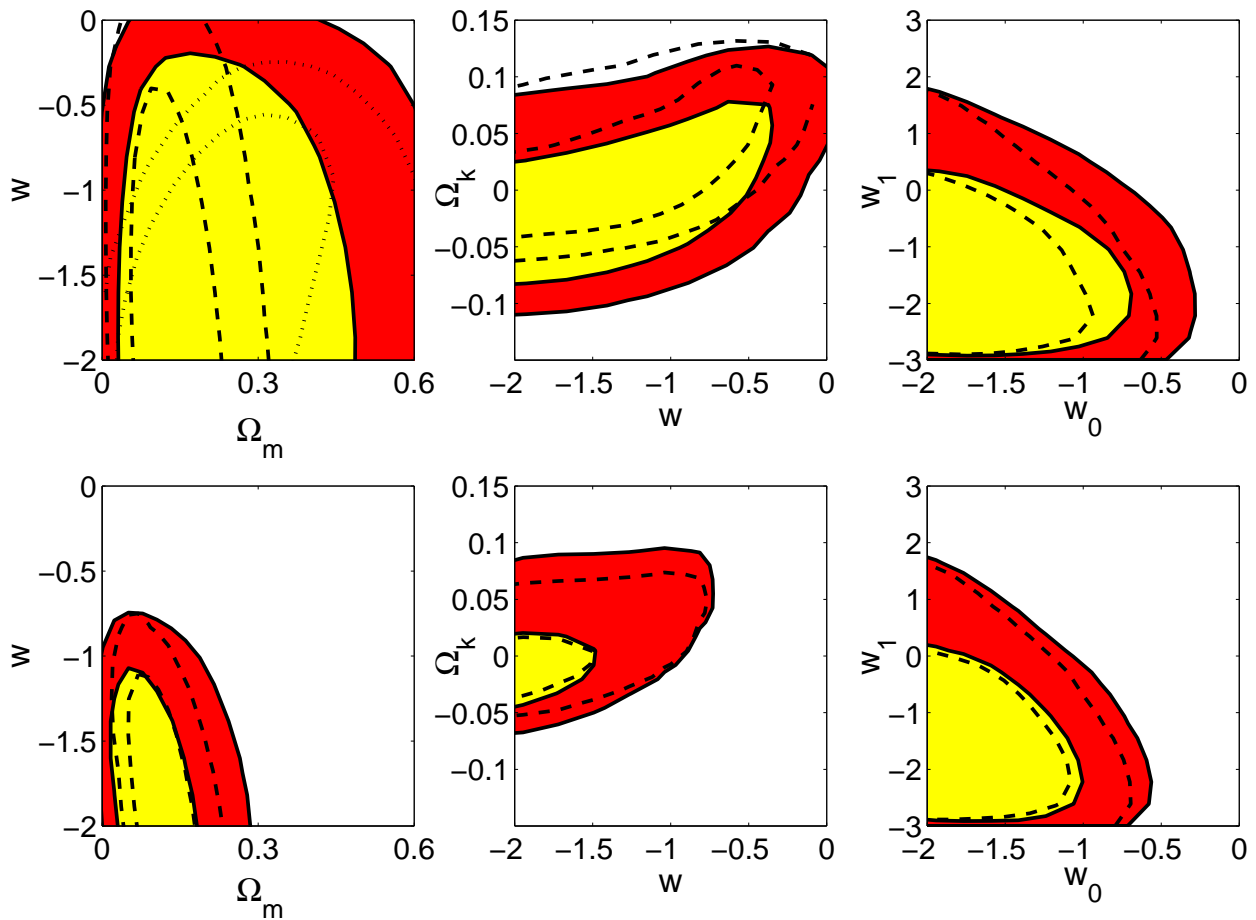}}
\caption{Marginalized $1$ and $2\sigma$ likelihood contours from WMAP
  extrema (upper panels) and in combination with BAO
  (lower panels). Dashed lines correspond to contours inferred under
  HST prior. The dotted lines in the upper left panel correspond to
  limits inferred assuming $\Omega_b=0.023$ and $n_s=0.96$.}
\label{fig3}
\end{figure*}

\section{Parameter Inference}\label{Cosmo}
We perform a Markov Chain Monte Carlo (MCMC) likelihood
analysis to derive cosmological parameter constraints using
the measurements of the WMAP extrema discussed in the previous
section. Again we account for the shift corrections by evaluating 
the model prediction for $l_m$ using Eq.~(\ref{lm}), 
with the displacements $\varphi_m$ parametrized as in \cite{Doran}.
We compute the recombination redshift $z_*$ using the fitting
formulae provided in \cite{HuSu}. Cosmological constraints derived
from the location of the CMB peaks have been presented in previous
works (e.g. \cite{Corasaniti,Doran2,Sen}). Here our aim is to derive 
bounds on dark energy which are independent of Supernova Ia
data and rely only on the cosmic distance information encoded in
the angular scale of the sound horizon as inferred from the multipole
position of the WMAP peaks and dips, and BAO measurements.

First we consider flat models with dark energy parametrized
by a constant equation of state $w$. We then test the stability of
the inferred constraints by extending the analysis to models 
with non-vanishing curvature, $\Omega_k\ne0$. 
We also consider flat dark energy models 
with a time varying equation of state 
parametrized as $w=w_0+w_1(1-a)$ (CPL) \cite{Polarski,Linder}. 
We want to remark that for models with $w_1\gg1$, 
the dark energy density can be non-negligible at early times. 
Therefore in order to consistently account for the shifts 
induced on the location of the CMB peaks and dips, we compute for each
model in the chain the corresponding value of $\Omega_{DE}^{dec}$ so
as to include its value in the shifts fitting formulae.

The credible intervals on the parameters of interest
are inferred after marginalizing over $h$, $\Omega_bh^2$
and $n_s$ respectively. We let them vary in the following intervals: 
$0.40<h<1.00$, $0.020<\Omega_bh^2<0.024$ and $0.94<n_s<1.10$.
Marginalizing over these parameters is necessary due to the parameter degeneracies 
in $r_K$, $r_s$ and to properly account for the shift corrections $\varphi_m$.

As complementary dataset we use the cosmic distance 
as inferred from the BAO in the SDSS and 2dF surveys \cite{Percival}.
These measurements consists of the ratio $r_s(z_*)/D_V(z)$, 
where $D_V(z)$ is a distance measure given by
\begin{equation}
D_V(z)=\left[(1+z)^2D_A(z)c z/H(z)\right]^{1/3},
\end{equation}
with $D_A(z)=r_K(z)/(1+z)$ the angular diameter distance at $z$. 
In particular Percival et al.~\cite{Percival} have found, 
$D_V(0.35)/D_V(0.2)=1.812\pm0.060$. 

In order to reduce the degeneracy with the Hubble parameter we also
infer constraints assuming a Gaussian HST prior $h=0.72\pm0.08$
\cite{HST}. In Figure~\ref{fig3} we plot the marginalized $1$ and
$2\sigma$ contours in the $\Omega_m$-$w$, $w$-$\Omega_K$ and
$w_0$-$w_1$ respectively. The upper panels correspond 
to constraints inferred from WMAP extrema alone, while the lower 
panels include the BAO data. Dashed contours are inferred under 
the HST prior. To be conservative we only quote 
marginalized $2\sigma$ limits. We now discuss these results 
in more detail.

\subsection{Limits from CMB peaks and dips}
\begin{figure}[t]
\centerline{
\includegraphics[scale=0.35]{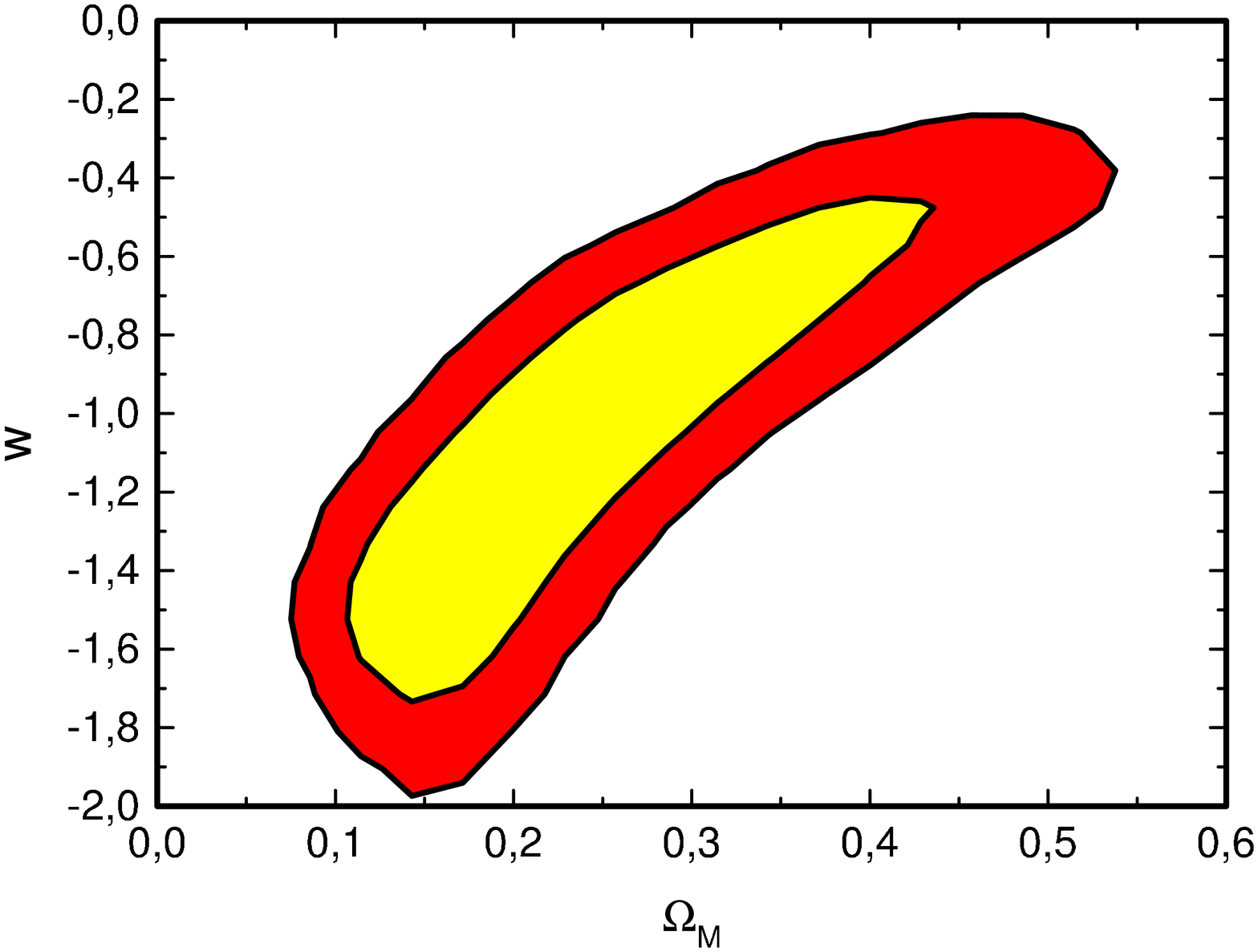}}
\caption{Marginalized $1$ and $2\sigma$ likelihood contours inferred from
the full WMAP-3yrs spectra.}
\label{fig4}
\end{figure}

As it can be seen in Figure~\ref{fig3} (upper left panel)
the CMB extrema alone poorly constrain the $\Omega_m-w$ plane.
In particular the $1$ and $2\sigma$ regions are larger than 
those obtained from the WMAP analysis \cite{Spergel}. 
This is because due to the late ISW effect more information 
about dark energy is contained in the full CMB spectrum 
than just in the distance to the last scattering surface
as encoded in the position of the CMB peaks and dips. Besides several
degeneracies with other parameters are strongly reduced. A
direct consequence of this is that our limits on $w$ are 
unbounded from below. After marginalizing 
over all parameters we find $\Omega_m=0.29\pm^{0.41}_{0.23}$ 
and $w<-0.18$ at $2\sigma$. A model with $\Omega_m=1$ 
is consistent at $95\%$ confidence level with the 
location of the WMAP extrema provided that $h\approx 0.42$. This is
in agreement with the results presented in \cite{Sarkar}. 
On the other hand imposing an
HST prior (dash contours) reduce the degeneracy in the $\Omega_m$-$w$
plane, and the marginalized $2\sigma$ limits are $\Omega_m=0.16\pm^{0.15}_{0.11}$ 
and $w<-0.25$ respectively. The upper limit on $w$ 
improves if a strong prior on $\Omega_b h^2$ 
and $n_s$ is assumed (dotted contours in the upper left panel). 
As an example imposing $\Omega_b h^2=0.0223$ and $n_s=0.96$, 
we find $w<-0.65$ at $2\sigma$. Indeed using the analysis of the full CMB power
spectrum provides better constraints. For instance in Fig.~\ref{fig4}
we plot the $1$ and $2\sigma$ contours inferred from a MCMC likelihood
analysis of the WMAP-3yrs spectra in combination with the HST prior. The limits are more 
stringent than in the previous case. This is because the amplitude of the first peak as well as 
the relative amplitude of the other peaks are particularly sensitive to 
$\Omega_m$, $\Omega_b$ and $h$. Hence degeneracies contributing to the uncertainties in 
the $\Omega_m-w$ plane are further reduced. As mentioned before, a robust dark energy parameter inference
needs the analysis of the full CMB spectrum. However in the case one aims
to infer constraints from other datasets such as SN Ia or BAO and include CMB information
in a rapid and simple manner, the position of the CMB extrema provides a very efficient tool.
In fact while the CMB power spectrum analysis requires the solution of the Boltzmann equation for 
a given cosmological model, the evaluation of the position of the CMB peaks and dips only 
is a semi-analytical computation. As an example
running publicly available Boltzamann codes \cite{Cmbcode} on a CPU at $2.3 {\rm GHz}$ requires about
one minute to compute the spectra of a single model, and even using an MCMC sampling the overall likelihood
analysis still require about one hour to reach full convergence of the MCMC chains, while using the CMB 
extrema only takes few minutes.

In Fig.~\ref{fig3} (central upper panel) we extend our analysis of the CMB peaks and dips
to non-flat models. Allowing for a non-vanishing curvature increases the geometric degeneracy 
and consequently leads to larger uncertainties in $w$. 
For instance the $2\sigma$ marginalized constraints are $w<-0.34$ 
and $\Omega_{K}=-0.01\pm{0.05}$ respectively, and 
do not improve significantly under the HST prior.

The position of the CMB peaks and dips alone 
does not provide any insight on the time variation of dark energy. 
As it can be seen in Fig.~\ref{fig3} (right upper panel) the contours
in the $w_0$-$w_1$ plane are spread over a large range of values.
After marginalizing we find $w_0<-0.55$ and $w_1<1.68$ at $2\sigma$.
It is worth mentioning that for increasing values of $w_1$, dark energy
becomes dominant at earlier times. In such a case the presence of a non-negligible dark energy 
density at recombination modifies the position of the CMB peaks and dips 
primarily through its effect on the size of the sound horizon at decoupling. 
Therefore the location of the CMB extrema (after having accouted for the relative shifts) 
can put an upper bound on the time evolution of the equation of state at high redshifts (i.e. $w_1$). 
Our analysis shows that in order to be consistent with the observed peak structure, 
large positive values of $w_1\gg1$ are excluded (see also Section~\ref{sndata}). 
This is consistent with the fact that the analysis of the full CMB spectrum limits the amount
of dark energy density at recombination to be less than $10\%$ (otherwise it would strongly affect
the amplitude and location of the CMB Doppler oscillations), hence providing a stringent upper
bounds on the value of the dark energy equation of state at early time (see \cite{Caldwell04,Coras04}).
In contrast models with large negative values of $w_1<0$ leave no imprint at high redshifts, 
since in this case the dark energy density rapidly decreases for $z>0$. Consequently
the likelihood remains unbounded in this region of the parameter
space.

\subsection{Combined constraints from CMB extrema and BAO}
The baryon acoustic oscillations in the galaxy power spectrum
provide a cosmic distance test at low redshifts. Therefore in
combination with CMB measurements they can significantly reduce 
the cosmological parameter degeneracies. 
In Fig.~\ref{fig3} (lower left panel) we plot the combined 
$1$ and $2\sigma$ contours in the $\Omega_m$-$w$
plane. At $95\%$ confidence level we find 
$\Omega_m=0.12\pm{0.12}$ and $w<-1.10$ respectively. 
Imposing the HST prior further constraints the dark energy equation
of state, $w<-1.14$. These results are compatible with those found in
\cite{Percival}. A model with $\Omega_m=1$ is now excluded 
with high confidence level since the combination of CMB 
extrema and BAO constrain the Hubble parameter
in the range $h=0.71\pm0.20$ at $2\sigma$ (see also \cite{Sarkar}).
Interestingly the $\Lambda$CDM case ($w=-1$) 
appears to be on the edge of the $2\sigma$
limit, hence favoring non-standard dark energy models.
Indeed unaccounted systematics effects in the BAO data can be
responsible for such super-negative values of $w$. On the other hand
if confirmed this would provide evidence for 
an exotic phantom dark energy component \cite{Caldwell} 
or interpreted as the cosmological signature 
of a dark sector interactions (e.g. \cite{Das}).

The credible regions for non-flat models are shown in Fig.~\ref{fig3}
(central lower panel). In this case we find $\Omega_K=-0.011\pm{0.064}$ 
and $w<-0.46$ at $2\sigma$. These bounds do not change significantly 
under the HST prior. In Fig.~\ref{fig3} (lower right panel) 
we plot the $1$ and $2\sigma$ contours in the $w_0$-$w_1$ plane. 
Also in this case the bounds on a time varying dark energy equation of
state remain large. For instance we find 
the marginalized $2\sigma$ limits to be $w_0<-0.74$ and $w_1<1.6$.
Necessarily inferring tighter bounds on $w_1$ will requires the
combination of several other datasets such as SN Ia 
luminosity distance measurements \cite{Zhao}, which is the topic 
of next Section.

\section{Constraints from SN Ia}\label{sndata}
\begin{figure}[t]
\includegraphics[scale=0.42]{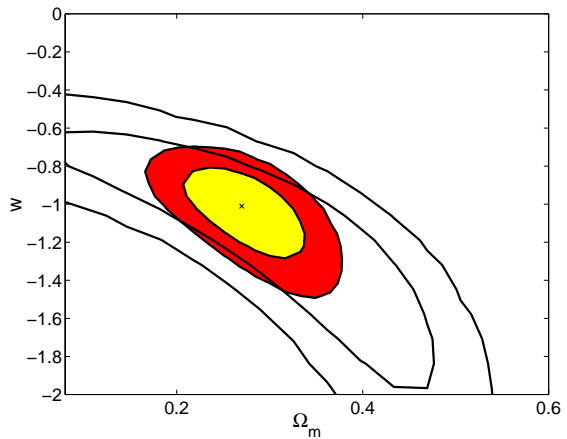}
\caption{Marginalized $1$ and $2\sigma$ contours in the $\Omega_m-w$ plane from
SNLS data (solid lines) and in combination with the location of the CMB extrema
(red and yellow shaded regions).}
\label{fig5}
\end{figure}

\begin{figure}[t]
\includegraphics[scale=0.42]{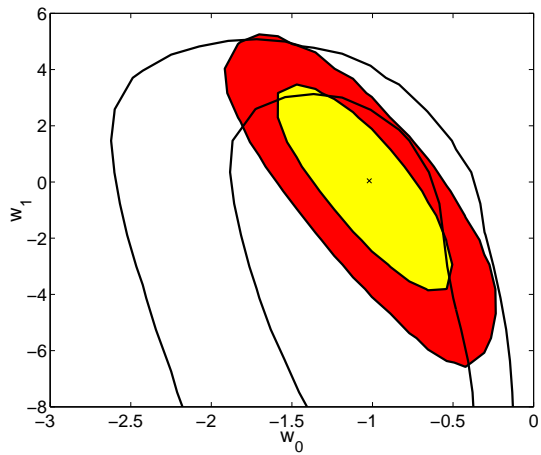}
\caption{As in Figure~\ref{fig5} in the $w_0-w_1$ plane.}
\label{fig6}
\end{figure}

 Here we want to compare the results derived in the previous Section
with limits inferred from luminosity distance measurements to SN Ia.
We use the SN dataset from the Supernova Legacy Survey (SNLS) \cite{Astier}, and
for simplicity we limit our analysis to flat models. The results are summarized
in Fig.~\ref{fig5} and Fig.~\ref{fig6} were we plot the $1$ and $2\sigma$ contours
in the $\Omega_m-w$ and $w_0-w_1$ planes respectively. The shades regions
correspond to limits inferred by combining the SN data with the location 
of the CMB extrema and assuming a hard prior on the baryon density and the scalar
spectral index, $\Omega_b=0.023$ and $n_s=0.96$ respectively. We have verified that 
the constraints do not change significantly assuming different prior parameter values.

Let first focus on Fig.~\ref{fig5}. We can see that
the degeneracy line in the $\Omega_m-w$ plane is almost orthogonal to that probed by CMB and BAO, 
and indeed using the SN data requires external information to extract tighter constraints on dark energy. 
A common procedure is to assume a Gaussian prior on $\Omega_m$ consistently with the parameter inference
from CMB and large scalar structure measurements, or alternatively to combine the SN analysis 
with BAO or the CMB shift parameter. Here we derive limits by combining the SN data with
the position of the CMB peaks and dips. This breaks the parameter degeneracy, thus
providing smaller ``credible'' contours (shaded contours). In particular after marginalizing, we find 
$\Omega_m=0.24\pm0.11$ and $w=-1.01\pm0.29$ at $2\sigma$ respectively. We can notice that these 
limits are only marginally consistent with those inferred using BAO in the previous Section, thus
indicating a potential discrepancy between the BAO measurements obtained in \cite{Percival} 
and the SNLS data \cite{Astier}. 

Let now consider the case of a time varying equation of state. It is obvious that the 
parameter degeneracy between the matter density and the dark energy equation of state 
is increased when additional equation of state parameters which accounts for a possible
redshift dependence are included in the data analysis. This can be clearly seen in Fig.~\ref{fig6} 
were we plot the $1$ and $2\sigma$ contours in the $w_0-w_1$ plane. Nevertheless the SN data, differently
from the case of BAO data in combination with CMB extrema (see lower left panel in Fig.~\ref{fig3}),
constrain $w_0$ in a finite interval. This is because SN Ia observations by testing the luminosity
distance over a range of redshift where the universe evolves from a matter dominated expansion to one
driven by dark energy, are sensitive to at least one dark energy parameter (i.e. $w$ or $w_0$) 
\cite{DraganLinder}. In such a case adding external information breaks the internal degeneracy
and leads to finite bounds on both dark energy parameters. For instance including the position of the CMB
peaks and dips, the root-mean-square value and standard deviation for $w_0$ and $w_1$ 
derived from the MCMC chains are $w_0=-1.04\pm0.33$ and $w_1=-0.27\pm2.27$ respectively; 
the best fit model being $w_0=-1.02$ and $w_1=0.04$. These results are consistent with those 
from other analysis in the literature (see e.g. \cite{Wang}).


\section{Shift Parameter}\label{shift}
The geometric degeneracy of the CMB power spectrum
implies that different cosmological models will have
similar spectra if they have nearly identical matter
densities $\Omega_m h^2$ and $\Omega_b h^2$, primordial
spectrum of fluctuations and shift parameter
$R=\sqrt{\Omega_m H_0^2} r_K(z_*)$ \cite{Bond}. 
The authors of \cite{Wang} have suggested that since 
$l_a$ is nearly uncorrelated with $R$, then both parameters can be
used to further compress CMB information and combined with other measurements
in a friendly user manner. For minimal extension of the dark energy parameters
the inferred values of $R$ and $l_a$ do not significantly differ from those
inferred assuming the vanilla $\Lambda$CDM model \cite{Elgaroy,Wang}. 
Indeed differences may arise if additional parameters, such as the 
neutrino mass, the running of the scalar spectral index 
or tensor modes are considered \cite{Elgaroy}. 
We extend this analysis to other models. 
In particular by running a MCMC likelihood analysis of the full 
WMAP-3yrs spectra we infer constraints on $R$ and $l_a$ 
for models with an extra-background of relativistic particles
(characterized by the number of relativistic species $N_{eff}\neq 3$) 
\cite{bowen}, neutrino mass \cite{fogli}, a time varying equation of state
parametrized in the form of CPL, and a dark energy component with 
perturbations characterized by the sound speed $c_{DE}^2$. 
We also consider models with a running of the scalar spectral index, with
a non-vanishing tensor contribution (see e.g. \cite{kinney}) 
and, finally, with extra-features in the primordial spectrum 
due to a sharp step in the inflaton potential as in \cite{covi}.

\begin{table}[t]\footnotesize
\begin{center}
\begin{tabular}{c|c|c}
Model & $R$ & $l_a$\\
\hline
$\Lambda$CDM& $1.707\pm0.025$ &
$302.3\pm1.1$\\
\hline
$w$CDM ($c_{DE}^2=1$)& $1.710\pm0.029$ &
$302.3\pm1.1$\\
\hline
$w$CDM ($c_{DE}^2=0$)& $1.711\pm0.025$ &
$302.4\pm1.1$\\
\hline
$\Lambda$CDM $m_{\nu} > 0$& $1.769\pm0.040$ &
$306.7\pm2.1$\\
\hline
$\Lambda$CDM $N_{eff} \neq 3$& $1.714\pm0.025$ &
$304.4\pm2.5$\\
\hline
$\Lambda$CDM $\Omega_k \neq 0$& $1.714\pm0.024$ &
$302.5\pm1.1$\\
\hline
$w(z)$CDM CPL ($c_{DE}^2=1$)& $1.710\pm0.026$ &
$302.5\pm1.1$\\
\hline
$\Lambda$CDM $+$ tensor& $1.670\pm0.036$ &
$302.0\pm1.2$\\
\hline
$\Lambda$CDM $+$ running& $1.742\pm0.032$ &
$302.8\pm1.1$\\
\hline
$\Lambda$CDM $+$ running $+$ tensor& $1.708\pm0.039$ &
$302.8\pm1.2$\\
\hline
$\Lambda$CDM $+$ features & $1.708\pm0.028$ &
$302.2\pm1.1$\\
\hline

\end{tabular}
\caption{The $68$\% C.L. limits on the shift parameter $R$ and the
acoustic scale derived from the WMAP data. A top-hat age
prior $10$ Gyrs $<t_0<20$ Gyrs is assumed.}
\label{table2}
\end{center}
\end{table}

As we can see from Table~\ref{table2} the constraints on $R$ and 
$l_a$ are stable under minimal modifications of the dark energy 
model parameters, differences are smaller than few per cent including the
case of a clustered dark energy component ($c_{DE}^2=0$).
In contrast the confidence interval of $l_a$ is shifted by few per
cent in the $\Lambda$CDM model with the neutrino mass or 
an extra background of relativistic particles, while
the values of $R$ are slightly modified for 
a running of the primordial power spectrum or the contribution of
tensor modes. These results confirm previous analysis \cite{Elgaroy,Wang}.

Although the values of $R$ and $l_a$ are nearly the same 
for the dark energy models we have considered,
this should not be considered as an incentive to use these parameters without caution. 
For instance there is no specific reason as to why one should use the values of
$R$ and $l_a$ inferred from the vanilla $\Lambda$CDM, rather than
those obtained accounting for the neutrino mass. Consequently one may infer
slightly different bounds on the dark energy parameters depending
whether neutrinos are assumed to be massless or not.
Moreover the fact that WMAP data constrain $R$ and $l_a$ to be nearly the
same for simple dark energy models is because the effect of dark energy
on the epoch of matter-radiation equality and the 
evolution of the density perturbations remains marginal. 
This might not be the case for other models, such as those for which
the dark energy density is a non-negligible at early
times. Since this effect is not accounted for in the values of $R$ and
$l_a$ inferred from the vanilla $\Lambda$CDM, their use may lead to strongly
biased results. In contrast the location of the CMB extrema 
is applicable to this class of models as well \cite{Doran}. 
A similar consideration applies to inhomogeneous models in which the
late times dynamics and geometry departs from that of the standard FRW 
universe \cite{Larena}. 

The applicability to models of modified gravity, 
such as the DGP scenario \cite{DGP} deserves a separate comment. 
In these models not only the Hubble law differs from the 
standard $\Lambda$CDM, but also the evolution of the
density perturbations can be significantly different. Therefore
unless the evolution of the linear perturbations is understood well enough
as to allow for a precise calculation of the CMB and matter power
spectra, the use $R$ and $l_a$, or alternatively of 
the position of the CMB extrema or the distance 
measurements from BAO might expose to the risk 
of completely wrong results.

\section{Conclusions}\label{Conclusions}

The multipoles of the CMB extrema can be directly measured from the 
WMAP spectra and used to combine CMB information with other
cosmological datasets. Corrections to the location of the CMB peaks
and dips from pre-recombination effects need to be taken into account for an 
unbiased parameter inference. Here we have shown that the position of the first
peak as measured by WMAP-3yrs data is strongly displaced with the respect to
the actual location of the acoustic horizon at recombination. 
This displacement is caused by gravitational driving forces 
affecting the propagation of sound waves
before recombination. These effects are smaller on higher harmonics,
indicating the presence of a scale dependent phase
shift which becomes negligible on scales well inside the horizon.

We have performed a cosmological parameter inference using the 
position of the WMAP peaks and dips in combination with recent BAO measurements
and derived constraints on a constant dark energy equation of state 
under different model parameter assumptions.

The method we have presented here is alternative to using
the shift parameter $R$ and/or the multipole of the acoustic 
horizon at decoupling $l_a$. We have tested for potential model
dependencies of $R$ and $l_a$ by running a 
full CMB spectra likelihood analysis for different class of models. 
Indeed for simple dark energy models 
the inferred constraints on $R$ and $l_a$ do not differ from
those inferred assuming the vanilla $\Lambda$CDM. 
Nevertheless we have suggested caution in using these secondary
parameters as {\it data}, since hidden assumptions may lead to biased results
particularly when testing models which greatly depart from the
$\Lambda$CDM cosmology. 

Indeed we do advocate the use of the full CMB spectra,
particularly for constraining the properties of dark energy.
In fact more information on dark energy is encoded in the full CMB spectrum
than just in the distance to the last scattering surface. Nevertheless
we think that using the location of the CMB extrema
provide a fast and self-consistent approach for combining in a 
friendly user way the CMB information with complementary cosmological data.

\section*{Acknowledgments} 
PSC is grateful to the Aspen Center for Physics for the hospitality 
during the ``Supernovae as Cosmological Distance Indicators'' Workshop 
where part of this work has been developped. We are particularly 
thankful to Jean-Michel Alimi, Laura Covi, Michael Doran, 
Malcom Fairbarn, Jan Hamann, Martin Kunz, Julien Larena, 
Eric Linder, Yun Wang and Martin White for 
discussions, suggestions and help.
We acknowledge the use of CosmoMC \cite{CosmoMC} for the 
analysis of the MCMC chains.

\end{document}